\documentstyle[floats,aps,epsf,prl]{revtex}
\begin{document}
\draft
\preprint{}
%
% [2]
\twocolumn[\hsize\textwidth\columnwidth\hsize\csname
@twocolumnfalse\endcsname
\title{Dynamic Local Distortions in Ferroelectrics}
\author{Henry Krakauer, Rici Yu,\cite{*} and Cheng-Zhang Wang}
\address{Department of Physics, College of William and Mary\\
Williamsburg, Virginia 23187--8795}
\author{Karin M. Rabe and Umesh V. Waghmare\cite{**}}
\address{Yale University, Department of Applied Physics, P. O. Box 208284\\
New Haven, Connecticut, 06520-8284}
\date{October 9, 1997}
\maketitle
\begin{abstract}

  Molecular-dynamics simulations of KNbO$_3$ reveal preformed dynamic
  chain-like structures, present even in the paraelectric phase, that
  are related to the softening of phonon branches over large regions
  of the Brillouin zone.  The phase sequence of ferroelectric
  transitions is correctly reproduced, showing that the
  first-principles effective Hamiltonian used in the simulations
  captures the essential behavior of the microscopic fluctuations
  driving the transitions.  Real space chains provide a framework for
  understanding both the displacive and order-disorder characteristics
  of these phase transitions.

\vskip 0.2truein\noindent {\sl Submitted to Physical Review Letters}
\end{abstract}
\vskip1pc]

There are many experimental indications in perovskite ferroelectrics
that the actual atomic structure in some of the phases may be
significantly different locally than is indicated by the average
crystallographic structure deduced from elastic X-ray and neutron
scattering. Among the earliest of these, diffuse X-ray scattering
measurements of Comes {\it et al.},\cite{Comesa,Comesb} on KNbO$_3$
and BaTiO$_3$ revealed temperature-dependent streak patterns, which
were interpreted as evidence for randomly ordered static chains
(oriented along $[100]$ directions) of distorted primitive cells, even
in the paraelectric cubic phase. Other observations such as
quasi-elastic central peaks in neutron scattering \cite{Shapiro72} and
Raman spectroscopy \cite{Fontana91} above the phase transition
temperature are also indicative of preformed clusters of the low
temperature phase. More recently, pair distribution functions obtained
from neutron scattering measurements up to very high momentum
transfers \cite{Teslic96} and XAFS measurements \cite{Stern96}
indicate the presence of short-range order. First-principles
calculations provide a powerful tool for probing the structural
energetics of local distortions.  For KNbO$_3$, LAPW linear response
calculation of the zero-temperature phonon dispersion in the cubic
perovskite structure reveals its instability against the formation of
localized chain distortions.\cite{Yu95}  In this paper, we
explore the implications of these results for dynamical behavior at
nonzero temperature through the molecular dynamics (MD) simulation of
an effective Hamiltonian constructed from first-principles
calculations, establishing the presence and dynamic nature of
localized chain distortions above $T_c$.

While first-principles calculations have provided a great deal of
information, they are limited to zero temperature and to relatively
small simulation cells. The use of effective
Hamiltonians, $H_{\it eff}$, to extend the reach of the 
first-principles results has
proved to be a useful strategy for the quantitative analysis of
temperature-driven structural phase transitions in real
materials.\cite{Zhong94,Rabe95,Waghmare97} These Hamiltonians act in
the subspace of the full ionic configuration space which contains the
degrees of freedom relevant to the transition(s).  These include, in
particular, the ``soft mode,'' identified as the unstable mode of the
high-symmetry structure which freezes in to produce the low-symmetry
phase(s).  The coefficients in a Taylor expansion around the
high-symmetry structure of the Born-Oppenheimer surface in this
subspace are determined from first-principles total-energy and
linear-response results. Nonzero temperature
simulations using $H_{\it eff}$ quantitatively reproduce the 
structural transition properties of the full system.

We constructed $H_{\it eff}$ for KNbO$_3$ using the
lattice-Wannier-function (LWF) method.\cite{Rabe95} 
Full details of the construction are presented
elsewhere,\cite{Krakauer97} and we give only a brief description here.
The effective Hamiltonian subspace is defined using a basis of
localized and symmetrized atomic displacement patterns, called lattice
Wannier functions, which are constructed to reproduce the
first-principles unstable polar $\Gamma_{15}$ phonon as well as
unstable transverse optic phonon eigenvectors and frequencies at other
high-symmetry points in the BZ \cite{Yu95}.  For KNbO$_3$,
the subspace is spanned by one vector degree of freedom per unit cell, 
$\xi_{i \alpha}$, representing the LWF coordinates,
where $i$ = unit cell index and $\alpha = x, y, z$.
We include as additional degrees of freedom the homogeneous
strain tensor, which describes changes in the overall volume and shape of the  
simulation cell. 

In the LWF basis, the kinetic energy retains a simple diagonal form.
The potential energy is expressed as a Taylor expansion in the LWF
coordinates $\xi_{i \alpha}$ and can be organized as follows: 
\begin{eqnarray}U =& &U_{on-site} + U_{short-range} + U_{dipolar} \nonumber \\ 
       & &+~ U_{LWF-strain} + U_{elastic}.
\label{Ueff}
\end{eqnarray}
We include anharmonic terms only in the on-site interaction
$U_{on-site} = \kappa \xi_i^2 + \delta \xi_i^4 + \gamma
  (\xi_{ix}^2 \xi_{iy}^2 +c.p.),$
and in the lowest order coupling between LWF coordinates (on-site
quadratic) and homogeneous strain (linear) $U_{LWF-strain}$. Strain
coupling of this type has been shown to be crucial in obtaining the
experimentally observed sequence of ferroelectric phase transitions in
perovskite ferroelectrics.\cite{Zhong94,Waghmare97b}
The coefficients of the onsite anharmonic and 
strain coupling terms were  
determined by fitting to first-principles total energies, varying strain and  
amplitude of uniform LWF distortions in the $[100]$, $[110]$ and $[111]$  
directions. The interactions between LWF coordinates in different 
unit cells are included  
to quadratic order only, with the general form
$\sum_{i j \alpha \beta} J_{i j \alpha \beta }
  \xi_{i \alpha} \xi_{j \beta}$.
Beyond third neighbor, the $J_{i j \alpha \beta }$ are
parameterized as the interaction between
two dipoles $Z^*\vec \xi_i$ and $Z^*\vec \xi_j$, where $Z^*$
is the mode effective
charge for the unstable zone-center phonon,
screened by the electronic dielectric constant $\varepsilon_{\infty}$.
$Z^*$ and $\varepsilon_{\infty}$ are computed directly using
LAPW linear response,  while
the short-range $J_{i j \alpha \beta }$ are fit to the first-principles
dynamical-matrix elements.

Using $H_{\it eff}$, classical molecular dynamics simulations were
carried out for a $10 \times 10 \times 10$ simulation cell,
corresponding to 5000 atoms, with periodic boundary conditions;
the $\xi_{i \alpha}$'s in $H_{eff}$ are in units of $10a$, 
where $a = 4.016 \AA $ is the
lattice constant.  A variable cell shape formalism was used together
with Nos\'{e}-Hoover thermostats to equilibrate the MD runs at
constant temperature.\cite{Broughton} After equilibration, and prior
to computing the static and dynamic structure factors, the thermostats
were turned off and the cell shape and volume were kept fixed. Further
equilibration (constant-energy MD) generally caused the temperature to
change by about 5 K. After this last equilibration, MD runs of
typically 20000 time steps (each time step $\sim$ 1 femtosecond) were
performed. The static and dynamic structure factors and 
autocorrelation functions of the $\xi_{i \alpha}$'s were then
computed\cite{Schneider78} using data from every 10th time step.

The different structural phases and transition temperatures $T_c$ are
identified in the MD simulations by calculating the three components
of the order-parameter $S_{\alpha}$, defined as an average over
the LWF coordinates: $S_{\alpha} = (1/N) \sum_i
\xi_{i \alpha}$. For example, the time
average of all three components of the order parameter is zero at 400
K in KNbO$_3$, indicating that the system is in the cubic paraelectric
phase.  At 370 K, one of the components of the order parameter freezes out with
a non-zero average value of about 0.16, but the time
average of the other two components is still zero, indicating that the
system is in the tetragonal phase.  Subsequent freezing out of the
other components signals transitions to the orthorhombic and
rhombohedral phases. The MD $T_c$ values were calculated by us for KNbO$_3$ 
and, 
as a calibration of the method,  
for BaTiO$_3$, using the $H_{eff}$ constructed in Ref. \cite{Zhong94}.
The results, given in Table I, are an average of cooling and heating runs, 
and  we estimate the error in these numbers to be about 5 - 10 K. 

\begin{table}
\caption[t1]{Comparison of calculated and measured transition 
  temperatures (see text), between the rhombohedral (R), orthorhombic
  (O), tetragonal (T) and cubic (C) phases. Temperatures are in
  Kelvin.}

\begin{tabular}{lddd}
                  & R - O & O - T & T - C \\ \tableline
KNbO$_3$                &  &  &  \\
MD     & 210 & 260 & 370      \\
Exper.\tablenotemark[3] & 210 - 265 & 488 & 701      \\
\tableline
BaTiO$_3$               &  &  &  \\
MD     & 200 & 230 & 290      \\
MC\tablenotemark[1]     & 197 & 230 & 290      \\
Exper.\tablenotemark[2] & 183 & 278 & 403      \\
\end{tabular}
\label{table1}
\tablenotetext[1]{Ref.~\onlinecite{Zhong94}.}
\tablenotetext[2]{From Ref.~\onlinecite{Zhong94}.}
\tablenotetext[3]{See for example, M. D. Fontana, G. Metrat, 
J. L. Servoin and F. Gervais, {\it J. Phys. C} {\bf 16}, 483 (1984).}
\end{table}

The calculated $T_c$ for the cubic-tetragonal
transition is significantly underestimated in both materials, and the error
is considerably larger in KNbO$_3$ than in BaTiO$_3$ and 
PbTiO$_3$ \cite{Waghmare97}.   
$T_c$'s for the R-O and O-T are
in better agreement, with the R-O agreement being the best.  
Possible sources of this quantitative discrepancy include the use of the  
local density approximation, the neglect of higher-order coupling to 
degrees of  
freedom outside the effective Hamiltonian subspace, and the sensitivity of the
transition to residual inaccuracies in the parametrization of strain coupling.
In any case, the nontrivial phase sequence and the trend from BaTiO$_3$ to 
KNbO$_3$  
are correctly reproduced, suggesting that the effective Hamiltonian 
captures the  essential behavior of the microscopic fluctuations 
driving the transitions.

\begin{figure}
\epsfxsize=3.3 truein
\centerline{\epsfbox{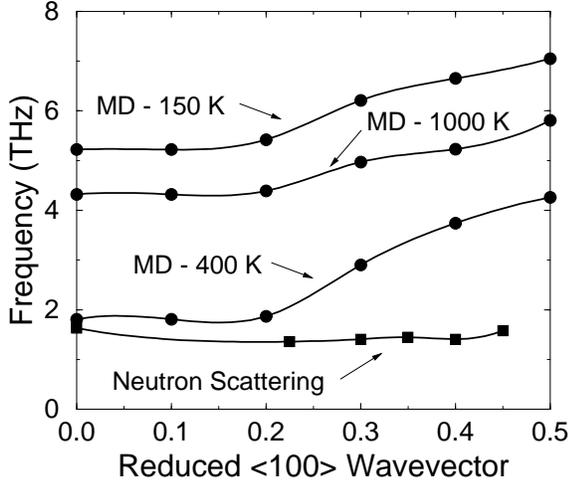}}
\caption{Comparison of the soft-mode dispersion along the [100] 
%                                      Ref. 15 = Holma96
  direction. Neutron scattering data from Ref.~[15] in the
  cubic phase of KNbO$_3$ at 798 K. In the MD simulations, the system
  is in the cubic phase at 400 and 1000 K and in the rhombohedral
  phase at 150 K.}
\label{fig.to.holma}
\end{figure}
Turning to the soft-mode vibrational frequencies, the temperature
dependence of the soft-mode dispersion in KNbO$_3$ is shown in
Fig.~\ref{fig.to.holma} and compared with the 798 K cubic phase
inelastic neutron scattering measurement of Holma and
Chen.~\cite{Holma96} The dispersion and temperature dependence were
determined from the dispersion of pronounced peaks in the Fourier
transformed autocorrelation function obtained from the MD simulations.
The experimental data is taken about 100 K above the cubic-tetragonal
phase transition. The 400 K MD results are 30 K above the calculated
transition temperature. The agreement is good below $q \sim$ 0.2, with
the theoretical TO branch showing greater dispersion for larger $q$.
This branch was unusually difficult to
measure for $q \geq$ 0,\cite{Holma96} however, 
the measured peak intensity being extremely low and the background
unusually high.

The most striking feature in
Fig. 1 is that the {\it entire} TO branch softens by about
2 THz on cooling from 1000 K to 400 K.  This behavior differs from the
usual soft-mode picture of a displacive phase transition, 
in which softening occurs only in the
vicinity of the wavevector associated with the structural phase
transition. It is consistent, however, with the previously obtained 
first-principles LAPW linear
response results.\cite{Yu95}
These revealed regions of instability in the BZ described by three
mutually perpendicular interpenetrating slabs centered at $\Gamma$,
perpendicular to the cubic axes, and extending to the BZ boundaries.
The first-principles unstable TO mode is seen in Ref.\cite{Yu95} to
also disperse slightly upwards in frequency
from $\Gamma$ to the $X$ point, consistent with what is found here. In the MD
simulations, however, the harmonically unstable first-principles TO branch is
anharmonically stabilized.

The fact that an entire phonon branch softens, {\it i.e.} the
softening occurs simultaneously over large regions of the BZ, suggests
that the analysis should focus on the temperature dependence of the structure
in real space
rather than in reciprocal space.  Tending more to a order-disorder picture, 
this also would naturally incorporate
the interpretation of the experimental observations of local
atomic structure, such as the streak patterns observed in diffuse
X-ray scattering in BaTiO$_3$ and KNbO$_3$.\cite{Comesb} Comes {\it et
  al.} invoked scattering from disordered finite-length static chains
of distorted primitive cells to explain this behavior, proposing an
empirical model in which there is sequential ordering of chains
directed along the three cubic axes. Thus, for example, in the cubic
to tetragonal transition, the ordering of the z-axis chains
corresponds to the disappearence of the incoherent scattering
(circular streaks in Fig. 6a of Ref.\cite{Comesb}) due to randomly oriented 
z-chains.  Subsequent
ordering of the x-axis and y-axis chains corresponds to entering the
orthorhombic and rhombohedral phases, with all the chains being
ordered in the ground state rhombohedral phase.  
The {\it ab initio} linear response
calculations of Yu and Krakauer\cite{Yu95} on KNbO$_3$ provided
theoretical support for the empirical chain model of
Comes {\it et al.} by showing the existence of BZ planar
instabilities. Furthermore, simulated diffuse elastic X-ray scattering
intensities calculated from MD simulations on BaTiO$_3$ reproduce the
observed position and sequential disappearance of the three families
(circles, vertical and horizontal) of streak patterns on cooling from
the cubic to rhombohedral phases.\cite{Krakauer97b}

\begin{figure}
\epsfxsize=3.3 truein
\centerline{\epsfbox{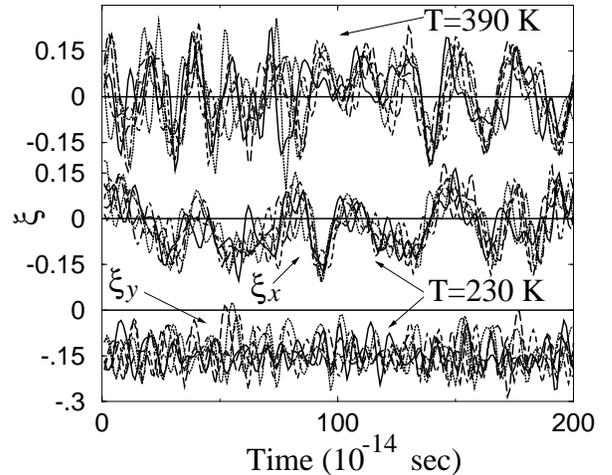}}
\caption{Time dependence of the LWF coordinate components
  \mbox{$\xi_{i x}$} and \mbox{$\xi_{i y}$} in the orthorhombic (230
  K) and cubic (390 K) phases of KNbO$_3$, 
  for six \mbox{$(i = 0 ... 5)$} adjacent
  primitive cells along an uncondensed chain in the $x$-direction (see
  text). $R_y$ and $R_z$ are condensed but $R_x = 0$. Chain-like
  correlations are evident in the correlated motion of the six
  LWF coordinates \mbox{$\xi_{i x}$} as a function of time, but
  not in the $y$ components, which oscillate randomly about the
  condensed value of -0.16.}
\label{fig.uxx230}
\end{figure}

\begin{figure}
\epsfxsize=3.3 truein
\centerline{\epsfbox{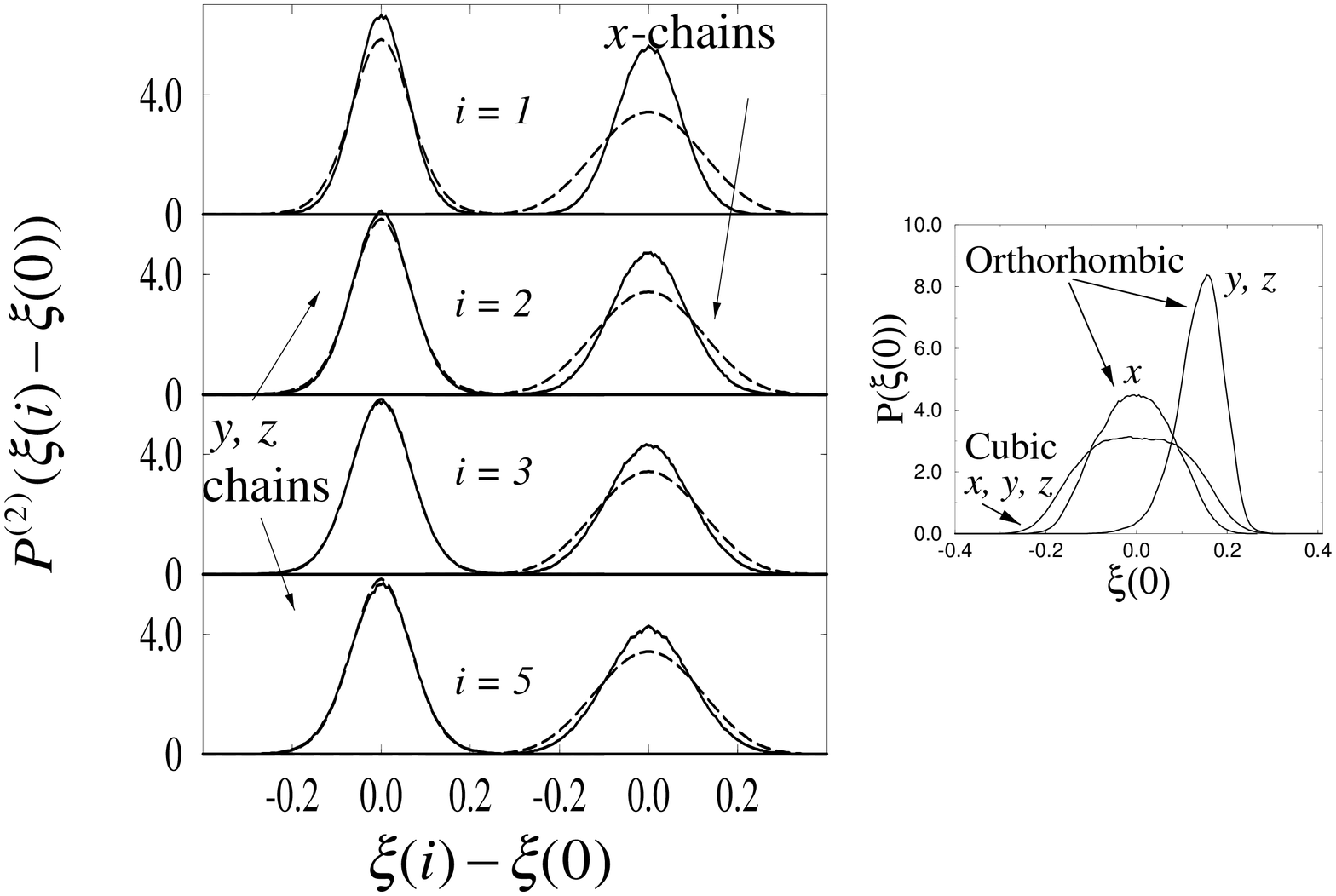}}
\caption{The solid lines are two-site equal-time probability distributions, 
  $P^{(2)}(\xi(i)-\xi(0))$, for i=1,2,3, and 5 along condensed $y$ and
  $z$ chains and for uncondensed $x$ chains in orthorhombic KNbO$_3$.
  In both cases, $\xi$ represents the longitudinal component of the
  LWF coordinate along the chain direction. The dashed lines
  are obtained using the one-site probability distribution, assuming
  no correlation between the two sites. The inset shows the one site
  probability distributions, $P(\xi(0))$, for the $x$, 
  $y$, and $z$ components in the orthorhombic phase, as well as in the cubic
  phase.}
\label{probsite}
\end{figure}

Diffuse X-ray scattering cannot, however, distinguish
between static and dynamic chains. Using MD simulations, we can resolve this  
issue. Fig.~\ref{fig.uxx230} shows the time dependence
of the LWF coordinates in the orthorhombic (230 K) and cubic
(390 K) phases of KNbO$_3$, for six adjacent primitive cells along the
$x$-direction. In the cubic phase all three order parameters are zero.
In the orthorhombic phase the order parameters $S_y$ and $S_z$ are
condensed, but the average value of $S_x$ is zero, corresponding to
disordered uncondensed chains along the x-direction.  Chain-like
correlations are evident in the correlated motion of the longitudinal
LWF components as a function of time in this figure. (Due to the
use of periodic boundary conditions in the \mbox{$10 \times 10 \times
  10$} simulation cell, the largest nearest neighbor distance along a 
chain is 5$a$.) As expected from the
linear response calculations\cite{Yu95}, the $y$- and $z$-components
of the atoms in an $x$-chain should be uncorrelated, and this is
confirmed in Fig.~\ref{fig.uxx230}. Similarly if we examine the
longitudinal (or transverse) LWF components along condensed $y$-
or $z$-chains (not shown) there is no correlation.  Chain correlation
is still evident even as high as 1000 K in the cubic phase, but the
characteristic chain reversal frequency increases as a function of
temperature.  Fig.~\ref{probsite} gives a more quantitative account of
the correlated motions.

These results unambiguously show 1) the existence in real space of
chains and 2) the {\it dynamic} character of the chains. This agrees
with the conclusions of Holma {\it et al.}\cite{Holma95}, based on
their diffuse X-ray measurements, and with the empirical lattice dynamics 
model of H\"uller \cite{Huller}. The MD simulations show that the
chains are preformed well above the cubic-tetragonal phase transition
temperature. The chains are defined by rows of distorted primitive
cells oriented along the three cubic axes, with the atomic
displacements along the chain highly correlated with one other.
Displacements in different chains are uncorrelated at high
temperature, and the observed phase transitions correspond to the
sequential freezing or onset of coherence of families of chains along
the three cubic axes. The softening of entire branches of unstable TO
modes gives rise to these real space chains and provides a framework
for understanding both the displacive and order-disorder
characteristics of these phase transitions.

\acknowledgments

Work at William and Mary was supported by Office of Naval Research
grant N00014-97-0049. Work at Yale was supported by ONR grant
N00014-97-1-0047 and the Alfred P. Sloan Foundation.  Computations
were carried out at the Cornell Theory Center. We are also pleased to
acknowledge the assistance of J. Broughton with the molecular dynamics
algorithms.

\end{document}